\documentclass[12pt]{iopart}

\usepackage{iopams}
\usepackage{graphicx}
\usepackage{color}
\usepackage{mathrsfs}
\usepackage{bm}

\begin{document}

\title{Modelling the Lagrangian of magnetized plasmas  with low-frequency magnetic perturbations by a gyrokinetic Ampere-Poisson model}

\author{Shuangxi Zhang }

%\address{%1. Institute of Fluid Physics, China Academy of Engineering Physics, Mianyang 621999, China. \\
% Graduate School of Energy Science, Kyoto University, Uji, Kyoto 611-0011, Japan. }

\address{Graduate School of Energy Science, Kyoto University, Uji, Kyoto 611-0011, Japan. }

%\address{%1. Institute of Fluid Physics, China Academy of Engineering Physics, Mianyang 621999, China. \\
% Graduate School of Energy Science, Kyoto University, Uji, Kyoto 611-0011, Japan. }

%\address{%1. Institute of Fluid Physics, China Academy of Engineering Physics, Mianyang 621999, China. \\
% Graduate School of Energy Science, Kyoto University, Uji, Kyoto 611-0011, Japan. }

\ead{zshuangxi@gmail.com;zhang.shuangxi.3s@kyoto-u.ac.jp}

\date{<date>}

\begin{abstract}
Following the method in Ref.(\cite{2017shuangxi}), this paper introduces a fundamental Lagrangian 1-form on the particle's coordinates, which determines the dynamics of all ions and electrons of the magnetized plasma with low-frequency magnetic perturbations. An Ampere-Vlasov model is utilized to model this fundamental Lagrangian 1-form based on a kind of coarse-grained scheme. With the Cary-Littlejohn single-parameter Lie transform method, a new fundamental Lagrangian 1-form on the gyrocenter coordinates is derived through transforming the one on particle's coordinates. This new 1-form determines the dynamics of all ions and electrons on the gyrocenter coordinates. A new Ampere-Vlasov model totally defined on the gyrocenter coordinates is developed to model the new Lagrangian 1-form. By incorporating the electrostatic perturbation into the Lagrangian 1-form, we eventually derived an Ampere-Poisson-Vlasov model defined on the gyrocenter coordinates.
\end{abstract}

%, i.e. the ratio of Larmor radius to characteristic spatial length of equilibrium magnetic field, the amplitude of perturbation of magnetic field, the amplitude of perturbation of electric field, the spatial gradient of perturbation wave

%\pacs{2.25.Dg, 52.30.Gz, 52.25.Xz, 52.55.Fa}% PACS, the Physics and Astronomy

\maketitle

\section{Introduction}\label{sec1}

Due to the multiple time-scale character of the magnetized plasma, the gyrokinetic theory is a strong and effective tool to help the numerical simulation of the magnetized plasma by decoupling the degree of freedom of the most fast time scale, i.e., the gyrating motion around the magnetic field line by the charged particles, from the remaining degrees of freedom.

%In old gyrokinetic Vlasov-Poisson (GVP) model, as pointed in Ref.(\cite{2017shuangxi}), the trajectory equations and Vlasov equation are defined on the gyrocenter coordinates, while the Poisson's equation is defined on particle's coordinates. The transform of the electrostatic potential, i.e., $\phi(\mathbf{x})=\phi(\mathbf{X}+\bm{\rho}(\mathbf{X},\mu,\theta))$, between the old coordinates and the gyrocenter coordinates induced by the condition $\mathbf{x}=\mathbf{X}+\bm{\rho}(\mathbf{X},\mu,\theta)$ in fact is a multivalued mapping, since dgiven a $\mathbf{x}$, for different $\mu$, there is a different $\mathbf{X}$ to make the condition satisfied. A multivalued mapping is not appropriate for the numerical simulation and can't be overcome in the old GVP model.

Ref.(\cite{2017shuangxi}) introduces a fundamental Lagrangian 1-form which determines the dynamics of all ions and electrons of the magnetized plasma with electrostatic perturbations, and develops a new GVP model as the modeling of the new 1-form on the gyrocenter coordinates transformed from the fundamental Lagrangian 1-form on the particle's coordinates. In the new model, the trajectories, Vlasov equation and Poisson's equation are defined on the gyrocenter coordinates, so that compared with the old GVP model, the numerical application of the new model could reduce the numerical error and instabilities, and simulation time.

Ref.(\cite{2017shuangxi}) focuses only on the electrostatic perturbation, while the magnetic field is treated as the background field. This paper takes the magnetic perturbation into account, more specifically, the low-frequency magnetic perturbation (LFMP). The reason why the low-frqueny magnetic perturbation is considered is as follows.

With the Lorentz gauge, the d'Alembert equations of the scalar potential and the vector potential for the electromagnetic wave in plasma are
\begin{equation}\label{af63}
{\nabla ^2}\phi  - \frac{1}{{\sqrt {{\mu _p}{\epsilon_p}} }}\frac{{{\partial ^2}{\phi}}}{{\partial {t^2}}} = \frac{{ - \rho }}{{{\epsilon_p}}},
\end{equation}
\begin{equation}\label{af64}
{\nabla ^2}{\bf{A}} - \frac{1}{{\sqrt {{\mu _p}{\epsilon_p}} }}\frac{{{\partial ^2}{\bf{A}}}}{{\partial {t^2}}} =  - {\mu _p}{\bf{J}}.
\end{equation}
$\mu_p$ and $\epsilon_p$ are the magnetic permeability and electric permittivity of the magnetized plasma, respectively. In toroidal magnetic field configuration, usually the magnetic or electric perturbation generated by the plasma is of very large radial gradient and low oscillating frequency that the condition $k_ \bot ^2 \gg \frac{{{\omega ^2}}}{{\sqrt {{\mu _p}{\epsilon_p}} }}$ is satisfied. For such kind of perturbation, the time-derivative terms  can be ignored in Eqs.(\ref{af63}) and Eq.(\ref{af64}), which are then simplified to be static equations as Poisson's equation and Ampere's law, respectively.  This paper focuses on this kind of perturbations.

This paper utilizes the similar method in Ref.(\cite{2017shuangxi}) to treat LFMP. The remaining part is arranged as follows. Sec.(\ref{sec2}) introduces the fundamental Lagrangian 1-form with existed LFMP. This Lagrangian 1-form determines the dynamics of all electrons and ions. An Ampere-Vlasov model is developed to model this fundamental Lagrangian 1-form based on a coarse-graining scheme. In Sec.(\ref{sec3}), the fundamental Lagrangian 1-form on the gyrocenter coordinate with the second order approximation is obtained by utilizing the Cary-Littlejohn single-parameter Lie transform method\cite{1983cary}. In Sec.(\ref{sec4}), a new Ampere-Vlasov model is derived to model the new fundamental Lagrangian 1-form.
In Sec.(\ref{sec5}), the electrostatic perturbation is incorporated into the model to form an Ampere-Poisson-Vlasov model. The appendix is the derivation of the new fundamental Lagrangian 1-form up to the second order approximation.

\section{Modeling the fundamental Lagrangian 1-form on particle's coordinates by Vlasov-Ampere model}\label{sec2}

The  magnetic field can be divided into the externally imposed part and the plasma-generated part. The externally imposed equilibrium magnetic field felt by the charged particle located at particle's spatial coordinates $\mathbf{x}_{oj}$ is written as $\mathbf{A}_{E}(\mathbf{x}_{oj})$. The subscript $E$ denotes the externally imposed field. The plasma-generated magnetic field includes equilibrium and non-equilibrium part. In this paper, the non-equilibrium part only includes LFMP, not the electromagnetic perturbation.  The plasma-generated  low-frequency magnetic vector potential felt by the particle located at $\mathbf{x}_{oj}$ can be written as
\begin{equation}\label{ap1}
{{\bf{A}}_{p}}\left( {{{\bf{x}}_{oj}}} \right) = \sum\limits_{o \in i,e} {\sum\limits_h ' {\frac{{{\mu _0}}}{{4\pi }}\frac{{{q_o}{{\bf{v}}_{oh}}}}{{|{{\bf{x}}_{oj}} - {{\bf{x}}_{oh}}|}}} },
\end{equation}
where $'$ denotes that $\mathbf{x}_{oj}$ is removed from the summation.
$\mu_0$ is the vacuum magnetic permittivity. The velocity $\mathbf{v}_{oh}$ of the charged particle located at $\mathbf{x}_{oh}$ can be divided into three parts: the part parallel to the direction of the equilibrium magnetic field; the perpendicular part depending on gyroangle; the drift part perpendicular to the magnetic field but independent of the gyroangle, so the velocity can be written as
\begin{equation}\label{ap2}
{{\bf{v}}_{oh}} = {U_{oh}}{\bf{b}} + {v_{oh \bot }}{\hat {\bf{v}}_{oh \bot }} +{{\bf{v}}_{ohd }},
\end{equation}
with ${\widehat {\bf{v}}_{oh \bot }} \equiv {{\bf{e}}_1}\sin {\theta _{oh}} + {{\bf{e}}_2}\cos {\theta _{oh}}$ and ${{\bf{v}}_{ohd }}$ is the drift velocity.
The perpendicular part depending on gyroangle is the origin of the magnetic moment, which is associated with the magnetic dipole radiation. Since only LFMP is accounted for, the magnetic dipole radiation is ignored in this paper.

The electrostatic potential is temporarily ignored. The fundamental Lagrangian 1-form can be written as
\begin{equation}\label{af3}
\gamma  = \sum\limits_{o \in \{ i,e\} } {\sum\limits_j {\left[ {\left( {{q_o}\left( {{{\bf{A}}_{E}}\left( {{{\bf{x}}_{oj}}} \right) + {{\bf{A}}_{p}}\left( {{{\bf{x}}_{oj}}} \right)} \right) + {m_o}{{\bf{v}}_{oj}}} \right)\cdot d{{\bf{x}}_{oj}} - \frac{1}{2}{m_o}v_{oj}^2dt} \right]} }
\end{equation}
Since the magnetic vector potential ${{\bf{A}}_{p}}\left( {{{\bf{x}}_{oj}}} \right)$ felt by the charged particle located at $\mathbf{x}_{oj}$ is generated by many-particle interaction, Eq.(\ref{ap1}) is not practical for the application. The way to calculate the magnetic vector potential can turn to the following Ampere's law alternatively
\begin{equation}\label{af4}
\begin{array}{*{20}{l}}
{ - {\nabla ^2}\left( {{{\bf{A}}_E}\left( {{{\bf{x}}_{oj}}} \right) + {{\bf{A}}_p}\left( {{{\bf{x}}_{oj}}} \right)} \right)}\\
{ = \frac{{{\mu _0}}}{{4\pi }}{{\bf{J}}_{0E}}\left( {{{\bf{x}}_{oj}}} \right) + \frac{{{\mu _0}}}{{4\pi }}\sum\limits_{o \in \{ i,e\} } {\sum\limits_{h'} {{q_o}{{\bf{v}}_{oj}}\delta \left( {{{\bf{x}}_{oj}} - {{\bf{x}}_{oh}}} \right)} } ,}
\end{array}
\end{equation}
based on the structure of ${{\bf{A}}_p}\left( {{{\bf{x}}_{oj}}} \right)$ given in Eq.(\ref{ap1}).
${{\bf{J}}_{E}}\left( {{{\bf{x}}_{oj}}} \right)$ is the current density at $\mathbf{x}_{oj}$ to generates the external magnetic vector field ${{{\bf{A}}_{E}}\left( {{{\bf{x}}_{oj}}} \right)}$. The plasma-generated ${{{\bf{A}}_p}\left( {{{\bf{x}}_{oj}}} \right)}$ can be extracted out as
\begin{equation}\label{af5}
- {\nabla ^2}{{\bf{A}}_p}\left( {{{\bf{x}}_{oj}}} \right) = \frac{{{\mu _0}}}{{4\pi }}\sum\limits_{o \in \{ i,e\} } {\sum\limits_{h'} {{q_o}{{\bf{v}}_{oj}}\delta \left( {{{\bf{x}}_{oj}} - {{\bf{x}}_{oh}}} \right)} }.
\end{equation}
However, this formula can't be straightforwardly used due to the existence of too many particles and no necessity of the knowledge of the accurate spatial position of each particle. Eq.(\ref{af5}) is usually modelled by a coarse-graining scheme. The coarse-graining scheme to model Poisson's equation is already introduced in Ref.(\cite{2017shuangxi}). Based on the coarse-graining scheme, the discrete edition of Eq.(\ref{af5}) is given as
\begin{equation}\label{af6}
- {\nabla ^2}{{\bf{A}}_{pk}}\left( {{{\bf{x}}_k}} \right) = \frac{{{\mu _0}}}{{4\pi }}\left( {{{\bf{J}}_{ik}}\left( {{{\bf{x}}_k}} \right) + {{\bf{J}}_{ek}}\left( {{{\bf{x}}_k}} \right)} \right).
\end{equation}
The derivative of ${{\bf{A}}_{pk}}\left( {{{\bf{x}}_k}} \right)$ at ${{{\bf{x}}_k}}$ is given by the middle-point discrete derivative. The magnetic vector potential felt ${{\bf{A}}_p}\left( {{{\bf{x}}_{oj}}} \right)$ by the $j$th particle located at the $k$th cell can be approximated by the discrete one ${{\bf{A}}_{pk}}\left( {{{\bf{x}}_k}} \right)$. The difference between the two quantities is approximately written as
\begin{equation}\label{af7}
{{\bf{A}}_p}\left( {{{\bf{x}}_{oj}},t} \right) = {{\bf{A}}_p}_k({{\bf{x}}_k},t) + O({l_c}).
\end{equation}
If $l_c$ is small enough to be close to $l_d$,  ${{\bf{A}}_p}_k({{\bf{x}}_k},t)$ can be close to ${{\bf{A}}_p}\left( {{{\bf{x}}_{oj}},t} \right)$. Then by replacing the accurate potential ${{\bf{A}}_p}\left( {{{\bf{x}}_{oj}},t} \right)$ by the approximate one ${{\bf{A}}_p}_k({{\bf{x}}_k},t)$, the Lagrangian 1-form for the $j$th particle located at the $k$th cell can be extracted out from the fundamental Lagrangian 1-form given in Eq.(\ref{af3}) as
\begin{equation}\label{af8}
\gamma_{okj}  = \left( {{q_o}\left( {{{\bf{A}}_E}\left( {{{\bf{x}}_{oj}}} \right) + {{\bf{A}}_{pk}}\left( {{{\bf{x}}_k}} \right)} \right) + {m_o}{{\bf{v}}_{oj}}} \right)\cdot d{{\bf{x}}_{oj}} - \frac{1}{2}{m_o}v_{oj}^2dt.
\end{equation}

The particle's distribution is modeled by the Klimontovich distribution
\begin{equation}\label{af9}
{M_{ok}}\left( {\bf{z}} \right) = \sum\limits_j {\delta \left( {{\bf{x}} - {{\bf{x}}_{oj}}(t)} \right)\delta \left( {{\bf{v}} - {{\bf{v}}_{oj}}\left( t \right)} \right)},
\end{equation}
the ensemble average of which gives the Vlasov distribution $f_o(\mathbf{z})$.
The time evolution of  $f_o(\mathbf{z})$ is
\begin{equation}\label{af10}
\left( {\frac{\partial }{{\partial t}} + \frac{{d{\bf{x}}}}{{dt}}\cdot\nabla  + \frac{{d{\bf{v}}}}{{dt}}\cdot\frac{\partial }{{\partial {\bf{v}}}}} \right)f_{ok}\left(\mathbf{z} \right) = 0.
\end{equation}
The trajectory equations $\frac{\partial \mathbf{x}}{\partial t}$ and $\frac{\partial \mathbf{v}}{\partial t}$ are given by the Lagrangian 1-form in Eq.(\ref{af8}).

So far, Eqs.(\ref{af6},\ref{af8},\ref{af10}) constitute the Vlasov-Ampere model to model the fundamental Lagrangian 1-form in Eq.(\ref{af3}).

\section{The fundamental Lagrangian 1-form on the new coordinates}\label{sec3}

The fundamental Lagrangian 1-form for all ions and electrons on particle's coordinates is given by Eq.(\ref{af3}). In this section, a pullback transform is utilized to pull the 1-form back to a new one on gyrocenter coordinates with $\theta$ angle decoupled from the remaining degrees of freedom up to order $O(\varepsilon_i^2)$ and $O(\varepsilon_e^{2})$. Here, the ordering parameters are given as ${\varepsilon _i} \equiv \frac{1}{{q{L_0}}}\sqrt {\frac{{2{\mu _{i }}{B_0}}}{{{m_i}}}}$ and ${\varepsilon _e} \equiv \frac{1}{{e{L_0}}}\sqrt {\frac{{2{\mu _{e}}{B_0}}}{{{m_e}}}} $. Three assumptions given in Ref.(\cite{2017shuangxi}) are also adopted here. The first one says that only the guiding equilibrium magnetic field affects the particles' gyrating orbit, while the perturbations only imposes the influence on the motion of the guiding center. So the coordinate transform can be almost written as $\psi_M^{-1}: (\mathbf{x},\mathbf{v})\to (\mathbf{X},\mu,U,\theta)$ with $\mathbf{x}=\mathbf{X}+\bm{\rho}(\mu,\theta)$ and $\bm{\rho} \left( {\mathbf{X},\mu ,\theta } \right) \equiv  - \frac{1}{{{q_o}}}\sqrt {\frac{{2{m_o}{\mu }}}{{B\left( {\bf{X}} \right)}}} \left( { - {{\bf{e}}_1}\cos \theta  + {{\bf{e}}_2}\sin \theta } \right)$ being the Larmor radius vector only caused by the equilibrium guiding magnetic field. It's the first order approximation of the transform ${\bf{x}} = \exp \left( { - {{\bf{g}}^{\bf{X}}}\cdot\nabla } \right){\bf{X}}$ with ${{\bf{g}}^{\bf{X}}}{\rm{ \equiv }}-\bm{\rho}$.

The new fundamental Lagrangian 1-form on the new coordinates can be derived based on the Cary-Littlejohn single-parameter Lie transform method. Its formula is written as
\begin{equation}\label{af11}
\Gamma  = \exp \left( { - \sum\limits_j {\left( {{\varepsilon _i}{L_{{\bf{g}}_{ij}^{\bf{X}}}} + {\varepsilon _e}{L_{{\bf{g}}_{ej}^{\bf{X}}}}} \right)} } \right)\gamma \left( {\bf{Z}} \right),
\end{equation}
where $\gamma$ is given by Eq.(\ref{af3}). Here, $\varepsilon_i$ and $\varepsilon_e$ only denote the order of the terms adjacent to them. This notation will be used throughout the rest of this paper. In Eq.(\ref{af11}), the generators are given as $\mathbf{g}_{ij}^{\mathbf{X}}=-\bm{\rho}_i$ and $\mathbf{g}_{ej}^{\mathbf{X}}=-\bm{\rho}_e$. Their order are $O(\varepsilon_i)$ and $O(\varepsilon_e)$, respectively. Since each coordinate pair $(\mathbf{x}_{oj},\mathbf{v}_{oj})$ is independent from all others, the operators $L_{\mathbf{g}_{oj}^\mathbf{x}}$ for each $j$ commutes. Expanding Eq.(\ref{af11}) based on the order of $\varepsilon_o$ and $\varepsilon_e$, the eventual Lagrangian differential 1-form independent of gyroangle can be derived.

Before expanding Eq.(\ref{af11}), the plasma-generated $\mathbf{A}_{p}$ is divided into the equilibrium and non-equilibrium parts. The equilibrium part is assumed to be generated by particles denoted by the subscript $h_1$
\begin{equation}\label{af12}
{{\bf{A}}_{p0}}\left( {{{\bf{x}}_{j}}} \right) = \frac{{{\mu _0}}}{{4\pi }}\sum\limits_{o \in i,e} {\sum\limits_{{h_1}}' {\frac{{{q_o}{{\bf{v}}_{o{h_1}}}}}{{|{{\bf{x}}_{j}} - {{\bf{x}}_{o{h_1}}}|}}} }.
\end{equation}
The non-equilibrium part is assumed to be generated by particles denoted by the subscript $h_2$
\begin{equation}\label{af13}
{{\bf{A}}_{p1}}\left( {{{\bf{x}}_{j}}},t \right) = \frac{{{\mu _0}}}{{4\pi }}\sum\limits_{o \in i,e} {\sum\limits_{{h_2}}' {\frac{{{q_o}{{\bf{v}}_{o{h_2}}}}}{{|{{\bf{x}}_{j}} - {{\bf{x}}_{o{h_2}}}(t)|}}} }.
\end{equation}
The explicit dependence on time by ${{\bf{A}}_{p1}}\left( {{{\bf{x}}_{oj}}},t \right)$ is due to the dependence on time by all $\mathbf{x}_{oh_2}$.
With the guiding center transform, the velocity can be decomposed into three parts as shown by Eq.(\ref{ap2}). The gyrating part around the magnetic field line is ignored in this paper. The detailed procedure to derive the fundamental Lagrangian 1-form, for which the second order approximation is taken and the gyroangle for each particle is decoupled from the remaining degrees of freedom, is given by Appendix.(\ref{app1}). The formula for the eventual Lagrangian 1-form on the gyrocenter coordinates is given by Eq.(\ref{af39}).

\section{Modeling the fundamental Lagrangian 1-form on the new coordinates by Vlasov-Ampere model}\label{sec4}

On the new coordinates, the plasma-generated magnetic vector potential felt by the $j$th particle is given by the formula in Eq.(\ref{af19}) and Eq.(\ref{af20}) for the equilibrium and perturbed part, respectively. They are of the same structure with that in Eq.(\ref{ap1}). The way to calculate these potentials can  be replaced by the Ampere's law plus a boundary condition. The Ampere's law on the new coordinates is
\begin{equation}\label{af40}
- \nabla _{oj}^2{{\bf{A}}_p}\left( {{{\bf{X}}_{oj}},t} \right) = \frac{{{\mu _0}}}{{4\pi }}\sum\limits_{o \in \{ i,e\} } {\sum\limits_h {{q_o}\left( {{U_{oh}}{\bf{b}}(\mathbf{X}_{oh}) + {{\bf{v}}_{ohd}}} \right)\delta \left( {{{\bf{X}}_{oj}} - {{\bf{X}}_{oh}}} \right)} }.
\end{equation}
Here, the derivative of $ {{\bf{v}}_{ohd}}$ is neglected.
The spatial length scale of ${{\bf{A}}_{p0}}\left( {{{\bf{X}}_{oj}},t} \right)$ is of the scale of the device. Usually, the spatial length scale of ${{\bf{A}}_{p1}}\left( {{{\bf{X}}_{oj}},t} \right)$ denoted as $l_p$ is much longer than the mean distance between charged particles denoted as $l_d$. To determine  the electrostatic potential , it's not needed to know the specific position of each particle. Alternatively, a coarse graining scheme can be utilized to divide the spatial area occupied by the plasma into small cells, with its length scale denoted by $l_c$. It's required that $l_p\gg l_c \gg l_d$ holds. Then, Eq.(\ref{af40}) can be approximated by a discrete edition as
\begin{equation}\label{af41}
- {\nabla ^2}{{\bf{A}}_{pk}}\left( {{{\bf{X}}_k},t} \right) = \frac{{{\mu _0}}}{{4\pi }}\left( {{{\bf{J}}_{ik}}\left( {{{\bf{X}}_k},t} \right) + {{\bf{J}}_{ek}}\left( {{{\bf{X}}_k},t} \right)} \right).
\end{equation}
The current ${{{\bf{J}}_{ik}}\left( {{{\bf{X}}_k},t} \right)}$ and ${{{\bf{J}}_{ek}}\left( {{{\bf{X}}_k},t} \right)}$ are obtained by integrating the velocity over the Kimontonvich distribution on the new coordinates located within the $k$th cell. The Kimontonvich distribution on the gyrocenter coordinates is
\begin{equation}\label{af42}
\mathbb{M}{_{ok}}\left( {\bf{Z}} \right) = \sum\limits_j {\frac{{\delta \left( {{\bf{X}} - {{\bf{X}}_{oj}(t)}} \right)\delta \left( {\mu  - {\mu _{oj}}(t)} \right)\delta \left( {U - {U_{oj}}(t)} \right)}}{{B\left( {{{\bf{X}}_{oj}}}(t) \right)}}}.
\end{equation}
The current is given as
\begin{equation}\label{af43}
{{\bf{J}}_{ok}}\left( {{{\bf{X}}_k},t} \right) = q_o\sum\limits_j {\frac{{\delta \left( {{\bf{X}} - {{\bf{X}}_{oj}}(t)} \right)\delta \left( {\mu  - {\mu _{oj}}(t)} \right)\delta \left( {U - {U_{oj}}(t)} \right){U_{oj}}{\bf{b}}\left( {{{\bf{X}}_{oj}}} \right)}}{{B\left( {{{\bf{X}}_{oj}}(t)} \right)}}}.
\end{equation}
In Eq.(\ref{af43}), the current caused by the drift velocity is ignored.
The ensemble average of the Klimontonvich distribution gives the Vlasov Equation.
The evolution of the Vlasov equation is obtained by the Liouville's equation
\begin{equation}\label{af44}
\left( {\frac{\partial }{{\partial t}} + \frac{{d{{\bf{X}}}}}{{dt}}\cdot\nabla  + \frac{{d{U}}}{{dt}}\frac{\partial }{{\partial U}}} \right)F{_{ok}}(\mathbf{Z}) = 0.
\end{equation}

Given the discrete equation Eq.(\ref{af41}), the real magnetic vector potential felt by the $j$th particle located at the $k$th cell is approximated by the discrete edition ${{\bf{A}}_{pk}}\left( {{{\bf{X}}_k},t} \right)$. Then, the test Lagrangian 1-form which determines the dynamics of the $j$th particle located within the $k$th cell can be extracted out from the fundamental Lagrangian 1-form in Eq.(\ref{af39}) and given as
\begin{equation}\label{af45}
\begin{array}{l}
{\Gamma _{okj}} = \left( \begin{array}{l}
{q_o}{{\bf{A}}_E}\left( {{{\bf{X}}_{oj}}} \right) + {q_o}{{\bf{A}}_{p0}}\left( {{{\bf{X}}_{oj}}} \right)\\
 + {\Lambda _{ojk}} + {m_o}{U_{oj}}{\bf{b}}
\end{array} \right)\cdot d{{\bf{X}}_{oj}}\\
 + \frac{{{m_o}{\mu _{oj}}}}{{{q_o}}}d{\theta _{oj}} - \left( {{\mu _o}B\left( {{{\bf{X}}_{oj}}} \right) + \frac{{{m_o}U_{oj}^2}}{2}} \right)dt,
\end{array}
\end{equation}
with
\begin{equation}\label{af46}
{\Lambda _{ojk}} = {q_o}{{\bf{A}}_{p1k}}\left( {{{\bf{X}}_k},t} \right) + \rho _{oj}^2\nabla _{oj}^2{A_{p1\parallel k}}\left( {{{\bf{X}}_k},t} \right){\bf{b}}.
\end{equation}
Term $\rho _{oj}^2\nabla _{oj}^2{A_{p1\parallel k}}\left( {{{\bf{X}}_k},t} \right){\bf{b}}$ is the FLR term contributing to the trajectory equations.

So far, we obtained a discrete Vlasov-Ampere model comprised of Eqs.(\ref{af41}),(\ref{af44}) and (\ref{af45}) as discrete editions of Ampere's law, the Vlasov equation, and the test Lagrangian 1-form, respectively, to approximate the fundamental Lagrangian 1-form given by Eq.(\ref{af39}), which determines the dynamics of all particle on the new coordinates with the second order approximation.  By shrinking the cell's length scale $l_c$ to be small enough, the subscript $k$ can be removed from the three equations, which are rewritten as
\begin{equation}\label{af47}
- {\nabla ^2}{{\bf{A}}_p}\left( {{\bf{X}},t} \right) = \frac{{{\mu _0}}}{{4\pi }}\left( {{{\bf{J}}_i}\left( {{\bf{X}},t} \right) + {{\bf{J}}_e}\left( {{\bf{X}},t} \right)} \right),
\end{equation}
\begin{equation}\label{af48}
\left( {\frac{\partial }{{\partial t}} + \frac{{d{\bf{X}}}}{{dt}}\cdot\nabla  + \frac{{dU}}{{dt}}\frac{\partial }{{\partial U}}} \right){F_o}({\bf{Z}}) = 0,
\end{equation}
\begin{equation}\label{af49}
\begin{array}{*{20}{l}}
{{\Gamma _o} = \left( {\begin{array}{*{20}{l}}
{{q_o}{{\bf{A}}_E}\left( {\bf{X}} \right) + {q_o}{{\bf{A}}_{p0}}\left( {\bf{X}} \right)}\\
{ + {\Lambda _o} + {m_o}{U_o}{\bf{b}}}
\end{array}} \right)\cdot d{{\bf{X}}_{oj}}}\\
{ + \frac{{{m_o}{\mu _o}}}{{{q_o}}}d{\theta _o} - \left( {{\mu _o}B\left( {{{\bf{X}}_o}} \right) + \frac{{{m_o}U_o^2}}{2}} \right)dt,}
\end{array}
\end{equation}
with
\begin{equation}\label{af50}
{\Lambda _o} = {q_o}{{\bf{A}}_{p1}}\left( {{\bf{X}},t} \right) + \rho _o^2\nabla _o^2{A_{p1\parallel }}\left( {{\bf{X}},t} \right){\bf{b}},
\end{equation}
\begin{equation}\label{af51}
{{\bf{J}}_o}\left( {{\bf{X}},t} \right) = {q_o}\int {dUd\mu U{\bf{b}}{F_o}\left( {\bf{Z}} \right)}.
\end{equation}

\section{Incorporate the electrostatic perturbation into the model}\label{sec5}

Now, we incorporate the electrostatic perturbations into the fundamental Lagrangian 1-form in Eq.(\ref{af3}) to get a new one as
\begin{equation}\label{af52}
\gamma  = \sum\limits_{o \in \{ i,e\} } {\sum\limits_j {\left[ \begin{array}{l}
\left( {{q_o}\left( {{{\bf{A}}_E}\left( {{{\bf{x}}_{oj}}} \right) + {{\bf{A}}_p}\left( {{{\bf{x}}_{oj}}} \right)} \right) + {m_o}{{\bf{v}}_{oj}}} \right)\cdot d{{\bf{x}}_{oj}}\\
 - \left( {\frac{1}{2}{m_o}v_{oj}^2 + \frac{1}{2}{q_o}\phi \left( {{{\bf{x}}_{oj}},t} \right)} \right)dt
\end{array} \right]} }
\end{equation}
with
\begin{equation}\label{af53}
\phi \left( {\bf{x}},t \right) = \frac{1}{{4\pi {\epsilon_0}}}\sum'\limits_{j} {\left( {\frac{q}{{|{{\bf{x}}} - {{\bf{x}}_{ij}}(t)|}} - \frac{e}{{|{{\bf{x}}} - {{\bf{x}}_{ej}}(t)|}}} \right)}.
\end{equation}

The new Lagrangian 1-form on the new coordinate can be derived by transforming $\gamma$ in Eq.(\ref{af52}) using the Cary-Littlejohn single-parameter Lie transform theory. With the same procedure presented in Ref.(\cite{2017shuangxi}) and in the previous sections in this paper, the Ampere-Poisson-Vlasov model as a modeling of the new fundamental Lagrangian 1-form on the new coordinate can be derived as
\begin{equation}\label{af54}
\left( {\frac{\partial }{{\partial t}} + \frac{{d{\bf{X}}}}{{dt}}\cdot\nabla  + \frac{{dU}}{{dt}}\frac{\partial }{{\partial U}}} \right){F_o}({\bf{Z}}) = 0,
\end{equation}
\begin{equation}\label{af55}
\begin{array}{*{20}{l}}
{{\Gamma _o} = \left( {\begin{array}{*{20}{l}}
{{q_o}{{\bf{A}}_E}\left( {\bf{X}} \right) + {q_o}{{\bf{A}}_{p0}}\left( {\bf{X}} \right)}\\
{ + {\Lambda _o} + {m_o}{U_o}{\bf{b}}}
\end{array}} \right)\cdot d{{\bf{X}}_{oj}}}\\
{ + \frac{{{m_o}{\mu _o}}}{{{q_o}}}d{\theta _o} - \left( {{\mu _o}B\left( {\bf{X}} \right) + \frac{{{m_o}U_o^2}}{2} + {\Psi _o}} \right)dt,}
\end{array}
\end{equation}
\begin{equation}\label{af56}
- {\nabla ^2}{{\bf{A}}_p}\left( {{\bf{X}},t} \right) = \frac{{{\mu _0}}}{{4\pi }}\left( {{{\bf{J}}_i}\left( {{\bf{X}},t} \right) + {{\bf{J}}_e}\left( {{\bf{X}},t} \right)} \right),
\end{equation}
\begin{equation}\label{af57}
{\nabla ^2}\Phi ({\bf{X}},t) = \frac{1}{{{\epsilon_0}}}\left( {e{N_e}({\bf{X}},t) - q{N_i}({\bf{X}},t)} \right),
\end{equation}
with
\begin{equation}\label{af58}
{\Psi _o} = {q_o}\Phi \left( {{\bf{X}},t} \right) + {q_o}\frac{{\rho _o^2{\nabla ^2}\Phi \left( {{\bf{X}},t} \right)}}{2},
\end{equation}
, $\Lambda_o$ given in Eq.(\ref{af50}), and $\mathbf{J}_o$ given in Eq.(\ref{af51}).
The spatial coordinates of Eqs.(\ref{af54}-\ref{af57}) are defined on the rectangular geometry. For the toroidal magnetic field configuration, the spatial coordinates of Eqs.(\ref{af54}-\ref{af57}) can be changed to fit the geometry.

The trajectory equations are derived from Eq.(\ref{af55})
\begin{equation}\label{af59}
\dot{\mathbf{X}}_o= \frac{{{m_o}{U_o}{\bf{B}}_o^* + {\bf{b}} \times \nabla {H_o}}}{{{q_o}{\bf{b}}\cdot{\bf{B}}_o^*}},
\end{equation}
\begin{equation}\label{af60}
{{\dot U}_o} = \frac{{ - {\bf{B}}_o^*\cdot\nabla {H_o}}}{{{m_o}{\bf{b}}\cdot{\bf{B}}_o^*}},
\end{equation}
with
\begin{equation}\label{af61}
{\bf{B}}_o^* = \nabla  \times \left( {{{\bf{A}}_E} + {{\bf{A}}_{p0}} + {\Lambda _o} - {U_o}{\bf{b}}} \right),
\end{equation}
\begin{equation}\label{af62}
{H_o} = {\mu _o}B\left( {\bf{X}} \right) + \frac{{{m_o}U_o^2}}{2} + {\Psi _o}.
\end{equation}

\appendix

\section{Expanding the new fundamental Lagrangian 1-form up to the second}\label{app1}

To expanding Eq.(\ref{af11}), the following two equations are needed
\begin{equation}\label{af14}
\begin{array}{l}
{L_{{\bf{g}}_1^{\bf{x}}}}\left( {{\bf{f}}({\bf{Z}})\cdot d{\bf{X}}} \right) =  - {\bf{g}}_1^{\bf{x}} \times \nabla  \times {\bf{f}}\left( {\bf{Z}} \right)\cdot d{\bf{X}}\\
 - {\bf{g}}_1^{\bf{x}}\cdot\left( {{\partial _t}{\bf{f}}({\bf{Z}})dt + {\partial _\theta }{\bf{f}}({\bf{Z}})d\theta  + {\partial _\mu }{\bf{f}}({\bf{Z}})d\mu } \right) + dS,
\end{array}
\end{equation}
\begin{equation}\label{af15}
{L_{{\bf{g}}_1^{\bf{x}}}}\left( {h({\bf{Z}})dt} \right) = {\bf{g}}_1^{\bf{x}}\cdot\nabla h\left( {\bf{Z}} \right)dt+dS.
\end{equation}
Here, $\mathbf{f}(\mathbf{Z})$ and $h(\mathbf{Z})$ are any vector function and scalar function on the new coordinates, respectively.

\subsection{Deriving $\Gamma_0$}

Among the expansions, the following 1-form is denoted as $\Gamma_0$
\begin{equation}\label{af16}
\begin{array}{*{20}{l}}
{{\Gamma _0} = \gamma ({\bf{Z}})}\\
{ = \sum\limits_{o \in \{ i,e\} } {\sum\limits_j {\left[ {\begin{array}{*{20}{l}}
{\left( \begin{array}{l}
{q_o}\left( {{{\bf{A}}_E}\left( {{{\bf{X}}_{oj}}} \right) + {{\bf{A}}_{p0}}\left( {{{\bf{X}}_{oj}}} \right) + {{\bf{A}}_{p1}}\left( {{{\bf{X}}_{oj}},t} \right)} \right)\\
 + {m_o}{U_{oj}}{\bf{b}} + {m_o}{{\bf{v}}_{ojd}}
\end{array} \right)\cdot d{{\bf{X}}_{oj}}}\\
{ - \left( {m_o}{U_{oh}^2 + {\mu _{oj}}B\left( {{{\bf{X}}_{oj}}} \right) + {m_o}v_{ojd}^2} \right)dt}
\end{array}} \right]} } .}
\end{array}
\end{equation}
In Eq.(\ref{af16}), the summation of the kinetic energy
\begin{equation}\label{af22}
\sum\limits_{o \in \{ i,e\} } {\sum\limits_h {{{\bf{v}}_{oh}}\cdot{{\bf{v}}_{oh}}} }  = \sum\limits_{o \in \{ i,e\} } {\sum\limits_h {\left( {U_{oh}^2 + {\mu _{oj}}B\left( {{{\bf{X}}_{oj}}} \right) + v_{ojd}^2} \right)} }
\end{equation}
is derived based on the following identities
%\begin{equation}\label{af17}
%{\left. {\sum\limits_l {{{\hat {\bf{v}}}_{ol \bot }}\cdot{{\hat {\bf{v}}}_{ol \bot }}} } \right|_{\scriptstyle{{\bf{X}}_{ol}} = {{\bf{X}}_1}\hfill\atop
%\scriptstyle{v_{ol \bot }} = {v_ \bot }\hfill}} = 1,
%\end{equation}
\begin{equation}\label{af18}
{\left. {\sum\limits_l {\left( {{{{\bf{\hat v}}}_{ol \bot }}\cdot{{{\bf{\hat v}}}_{old}}} \right)} } \right|_{\scriptstyle{{\bf{X}}_{ol}} = {{\bf{X}}_1},{U_{ol}} = {U_1}\hfill\atop
\scriptstyle{v_{ol \bot }} = {v_{ \bot 1}},{{\bf{v}}_{old}} = {{\bf{v}}_{od1}}\hfill}} = 0
\end{equation}
Here, $\mathbf{X}_1,v_\perp,\mathbf{v}_{od}$ is a group of coordinates to characterize the gyrocenter.
Both Eqs.(\ref{af22}) and (\ref{af18}) are based on the homogeneous assumption of the distribution in the $\theta$ direction. To derive Eq.(\ref{af16}), the following identity is also used
\begin{equation}\label{af21}
{\left. {\sum\limits_l {\sqrt {\frac{{2B\left( {{{\bf{X}}_{ol}}} \right){\mu _{ol}}}}{{{m_o}}}} } {{\widehat {\bf{v}}}_{ol}}\cdot d{{\bf{X}}_{ol}}} \right|_{\scriptstyle{{\bf{X}}_{ol}} = {{\bf{X}}_1}\hfill\atop
{\scriptstyle{U_{ol}} = {U_1}\hfill\atop
\scriptstyle{\mu _{ol \bot }} = {\mu _1}\hfill}}} = 0
\end{equation}
with ${\widehat {\bf{v}}_{ol}} = {\bf{e}}\sin {\theta _{ol}} + {{\bf{e}}_2}\cos {\theta _{ol}}$.

Terms ${{{\bf{A}}_{p0}}\left( {{{\bf{X}}_{oj}}} \right)}$ and ${{{\bf{A}}_{p1}}\left( {{{\bf{X}}_{oj}},t} \right)}$ in Eq.(\ref{af16}) are
\begin{equation}\label{af19}
{{\bf{A}}_{p0}}\left( {{{\bf{X}}_{oj}}} \right) = \frac{{{\mu _0}}}{{4\pi }}\sum\limits_{o \in i,e} {\sum\limits_{{h_1}}' {\frac{{{q_o}\left( {{U_{o{h_1}}}{\bf{b}}(\mathbf{X}_{oh_1}) + {{\bf{v}}_{o{h_1}d}}} \right)}}{{|{{\bf{X}}_{oj}} - {{\bf{X}}_{o{h_1}}}|}}} },
\end{equation}
\begin{equation}\label{af20}
{{\bf{A}}_{p1}}\left( {{{\bf{X}}_{oj}},t} \right) = \frac{{{\mu _0}}}{{4\pi }}\sum\limits_{o \in i,e} {\sum\limits_{{h_2}}' {\frac{{{q_o}\left( {{U_{o{h_2}}}{\bf{b}}(\mathbf{X}_{oh_2}) + {{\bf{v}}_{o{h_2}d}}} \right)}}{{|{{\bf{X}}_{oj}} - {{\bf{X}}_{o{h_2}}}|}}} }.
\end{equation}

\subsection{Deriving $\Gamma_1$}

The next one is
\begin{equation}\label{af22}
\begin{array}{l}
{\Gamma _1} =  - \sum\limits_{o \in \{ i,e\} } {\sum\limits_j {{\varepsilon _o}{L_{{\bf{g}}_{oj}^{\bf{X}}}}\gamma \left( {\bf{Z}} \right)} } \\
 = \sum\limits_{o \in \{ i,e\} } {\sum\limits_j {\left[ { - {\varepsilon _o}{L_{{\bf{g}}_{oj}^{\bf{X}}}}\left( {{\gamma _{oj{\bf{X}}}}\cdot d{{\bf{X}}_{oj}}} \right) + {\varepsilon _o}{L_{{\bf{g}}_{oj}^{\bf{X}}}}\left( {{\gamma _{ojt}}dt} \right)} \right]} } .
\end{array}
\end{equation}
with
\begin{equation}\label{af23}
\begin{array}{*{20}{l}}
{{\gamma _{oj{\bf{X}}}}\cdot d{{\bf{X}}_{oj}}}\\
{ = \left( {\begin{array}{*{20}{l}}
{\frac{{{q_o}}}{{{\varepsilon _o}}}\left[ {{{\bf{A}}_E}\left( {{{\bf{X}}_{oj}}} \right) + {{\bf{A}}_{p0}}\left( {{{\bf{X}}_{oj}}} \right) + {{\bf{A}}_{p1}}\left( {{{\bf{X}}_{oj}},t} \right)} \right]}\\
{ + {m_o}{U_{oj}}{\bf{b}} + \sqrt {\frac{{2B({{\bf{X}}_{oj}}){\mu _{oj}}}}{{{m_o}}}} {{\widehat {\bf{v}}}_{oj \bot }} + {m_o}{{\bf{v}}_{ojd}}}
\end{array}} \right)\cdot d{{\bf{X}}_{oj}},}
\end{array}
\end{equation}
\begin{equation}\label{af24}
{\gamma _{ojt}}dt = \left( {m_o U_{oj}^2 + {\mu _{oj}}B\left( {{{\bf{X}}_{oj}}} \right) + m_o v_{oj \bot }^2} \right)dt.
\end{equation}

To simplify Eq.(\ref{af22}), the following identities based on the homogeneous assumption are needed
\begin{equation}\label{af25}
\begin{array}{l}
\sum\limits_j {{L_{{\bf{g}}_{oj}^{\bf{X}}}}\left( {\left[ {{{\bf{A}}_E}\left( {{{\bf{X}}_{oj}}} \right) + {{\bf{A}}_{p0}}\left( {{{\bf{X}}_{oj}}} \right) + {m_o}{U_{oj}}{\bf{b}}} \right]\cdot d{{\bf{X}}_{oj}}} \right)} \\
 =  - \sum\limits_j {{\bf{g}}_{oj}^{\bf{X}} \times \nabla  \times \left( {{{\bf{A}}_E}\left( {{{\bf{X}}_{oj}}} \right) + {{\bf{A}}_{p0}}\left( {{{\bf{X}}_{oj}}} \right) + {m_o}{U_{oj}}{\bf{b}}} \right)\cdot d{{\bf{X}}_{oj}}}  = 0
\end{array}
\end{equation}
\begin{equation}\label{af26}
\begin{array}{l}
\sum\limits_j {{L_{{\bf{g}}_{oj}^{\bf{X}}}}\left( {{{\bf{A}}_{p1}}\left( {{{\bf{X}}_{oj}},t} \right)\cdot d{{\bf{X}}_{oj}}} \right)} \\
 = \sum\limits_j { - {\bf{g}}_{oj}^{\bf{X}} \times \nabla  \times {{\bf{A}}_{p1}}\left( {{{\bf{X}}_{oj}}} \right)\cdot d{{\bf{X}}_{oj}} - {\bf{g}}_{oj}^{\bf{X}}\cdot{\partial _t}{{\bf{A}}_{p1}}\left( {{{\bf{X}}_{oj}},t} \right)dt}  = 0,
\end{array}
\end{equation}
where ${{\bf{A}}_{p1}}\left( {{{\bf{X}}_{oj}},t} \right)$ given by Eq.(\ref{af20}) doesn't include subscript $j$ and
\begin{equation}\label{af27}
{\partial _t}{{\bf{A}}_{p1}}\left( {{{\bf{X}}_{oj}},t} \right) = \sum'\limits_h {\frac{{d{{\bf{X}}_{oh}}}}{{dt}}} \cdot{\nabla _{oh}}{{\bf{A}}_{p1}}\left( {{{\bf{X}}_{oj}},t} \right),
\end{equation}
and
\begin{equation}\label{af28}
\sum\limits_j {{L_{{\bf{g}}_{oj}^{\bf{X}}}}\left( {{\gamma _t}\left( {\bf{Z}} \right)dt} \right) = 0},
\end{equation}
\begin{equation}\label{af29}
\begin{array}{*{20}{l}}
{{\varepsilon _o}\sum\limits_j {{L_{{\bf{g}}_{oj}^{\bf{X}}}}\left( {{v_{oj \bot }}{{\widehat {\bf{v}}}_{oj \bot }}\cdot d{{\bf{X}}_{oj}}} \right)} }\\
{ = {\varepsilon _o}\sum\limits_j {\left[ {\begin{array}{*{20}{l}}
{ - {\bf{g}}_{oj}^{\bf{X}} \times \nabla  \times \left( {{v_{oj \bot }}{{\widehat {\bf{v}}}_{oj \bot }}} \right)\cdot d{{\bf{X}}_{oj}}}\\
{ - {\bf{g}}_{oj}^{\bf{X}}\cdot{\partial _{{\theta _{oj}}}}\left( {{v_{oj \bot }}{{\widehat {\bf{v}}}_{oj \bot }}} \right)d{\theta _{oj}}}\\
{ - {\bf{g}}_{oj}^{\bf{X}}\cdot{\partial _{{\mu _{oj}}}}\left( {{v_{oj \bot }}{{\widehat {\bf{v}}}_{oj \bot }}} \right)d{\mu _{oj}}}
\end{array}} \right]} }\\
{ \approx {\varepsilon _o}\sum\limits_j {\frac{{2{m_o}{\mu _{oj}}}}{{{q_o}}}d{\theta _{oj}}} .}
\end{array}
\end{equation}
In Eq.(\ref{af29}), the $\mathbf{X}_{oj}$ components for all $j$ are neglected, since it's of order $O(\varepsilon_o)$, while the corresponding terms in the $\mathbf{X}_{oj}$ components in $\Gamma_0$ are of order $\varepsilon_o^{-1}$. On the other hand, this term can also be cancelled by introducing a generator $\mathbf{g}_{2oj}^{\mathbf{X}}$ of order $\varepsilon_{o}^2$.

Eventually, $\Gamma_1$ is simplified to be
\begin{equation}\label{af30}
{\Gamma _1} = {\varepsilon _o}\sum\limits_{o \in \{ i,e\} } {\sum\limits_j {\frac{{2{m_o}{\mu _{oj}}}}{{{q_o}}}d{\theta _{oj}}} }.
\end{equation}

\subsection{Deriving $\Gamma_2$}

$\Gamma_2$ is given by the following formula
\begin{equation}\label{af31}
{\Gamma _2} = \frac{1}{2}\sum\limits_{o,n \in \{ i,e\} } {\sum\limits_{j,h} {{\varepsilon _o}{\varepsilon _n}{L_{{\bf{g}}_{oj}^{\bf{X}}}}{L_{{\bf{g}}_{nh}^{\bf{X}}}}\gamma \left( {\bf{Z}} \right)}}.
\end{equation}
The following identities are needed
\begin{equation}\label{af32}
\begin{array}{l}
{\varepsilon _o}L_{{\bf{g}}_{oj}^{\bf{X}}}^2\left[ {\left( {{{\bf{A}}_E}\left( {{{\bf{X}}_{oj}}} \right) + {{\bf{A}}_{p0}}\left( {{{\bf{X}}_{oj}}} \right)} \right)\cdot d{{\bf{X}}_{oj}}} \right]\\
 = {\bf{g}}_{oj}^{\bf{X}} \times {\nabla _{oj}} \times \left( {{\bf{g}}_{oj}^{\bf{X}} \times {\nabla _{oj}} \times \left( {{{\bf{A}}_E}\left( {{{\bf{X}}_{oj}}} \right) + {{\bf{A}}_{p0}}\left( {{{\bf{X}}_{oj}}} \right)} \right)} \right)\cdot d{{\bf{X}}_{oj}}\\
 - {\varepsilon _o}{q_o}{\bf{g}}_{oj}^{\bf{X}}\cdot{\partial _{{\theta _{oj}}}}\left( {{\bf{g}}_{oj}^{\bf{X}} \times {\bf{B}}\left( {{{\bf{X}}_{oj}}} \right)} \right)d{\theta _{oj}}\\
 \approx  - {\varepsilon _o}\rho _0^2B\left( {{{\bf{X}}_{oj}}} \right) =  - {\varepsilon _o}\frac{{2{m_o}{\mu _{oj}}}}{{{q_0}}}d{\theta _{oj}}.
\end{array}
\end{equation}
\begin{equation}\label{af33}
\begin{array}{l}
{\varepsilon _o}L_{{\bf{g}}_{oj}^{\bf{X}}}^2\left[ {{{\bf{A}}_{p1}}\left( {{{\bf{X}}_{oj}},t} \right)\cdot d{{\bf{X}}_{oj}}} \right]\\
 = {\varepsilon _o}{\bf{g}}_{oj}^{\bf{X}} \times {\nabla _{oj}} \times \left( {{\bf{g}}_{oj}^{\bf{X}} \times {\nabla _{oj}} \times {{\bf{A}}_{p1}}\left( {{{\bf{X}}_{oj}},t} \right)} \right)\cdot d{{\bf{X}}_{oj}}\\
 - {\varepsilon _o}{\bf{g}}_{oj}^{\bf{X}}\cdot{\nabla _{oj}}\left( {{\bf{g}}_{oj}^{\bf{X}}\cdot{\partial _t}{{\bf{A}}_{p1}}\left( {{{\bf{X}}_{oj}},t} \right)} \right)\\
 \approx {\varepsilon _o}{\bf{g}}_{oj}^{\bf{X}} \times {\nabla _{oj}} \times \left( {{\bf{g}}_{oj}^{\bf{X}} \times {\nabla _{oj}} \times {{\bf{A}}_{p1}}\left( {{{\bf{X}}_{oj}},t} \right)} \right)\cdot d{{\bf{X}}_{oj}}
\end{array}
\end{equation}
The $``\approx"$ in Eq.(\ref{af33}) is derived by assuming the direction of ${{{\bf{A}}_{p1}}\left( {{{\bf{X}}_{oj}},t} \right)}$ mainly parallel to the direction of the equilibrium magnetic field and the gradient of ${{{\bf{A}}_{p1}}\left( {{{\bf{X}}_{oj}},t} \right)}$ mainly in the perpendicular direction. The last term in Eq.(\ref{af33}) can be simplified to be
\begin{equation}\label{af34}
\begin{array}{l}
{\bf{g}}_{oj}^{\bf{X}} \times {\nabla _{oj}} \times \left( {{\bf{g}}_{oj}^{\bf{X}} \times {\nabla _{oj}} \times {{\bf{A}}_{p1}}\left( {{{\bf{X}}_{oj}},t} \right)} \right)\\
 = {\bf{g}}_{oj}^{\bf{X}} \times {\nabla _{oj}} \times \left( {{\bf{g}}_{oj}^{\bf{X}} \times {{\bf{B}}_{p1}}\left( {{{\bf{X}}_{oj}},t} \right)} \right)\\
 = {\bf{g}}_{oj}^{\bf{X}} \times \left[ {{\bf{g}}_{oj}^{\bf{X}}\left( {\nabla \cdot{{\bf{B}}_{p1}}} \right) - {{\bf{B}}_{p1}}\left( {\nabla \cdot{\bf{g}}_{oj}^{\bf{X}}} \right) + \left( {{{\bf{B}}_{p1}}\cdot\nabla } \right){\bf{g}}_{oj}^{\bf{X}} - \left( {{\bf{g}}_{oj}^{\bf{X}}\cdot\nabla } \right){{\bf{B}}_{p1}}} \right]\\
 \approx  - {\bf{g}}_{oj}^{\bf{X}} \times \left( {{\bf{g}}_{oj}^{\bf{X}}\cdot\nabla } \right){{\bf{B}}_{p1}} =  - {\bf{g}}_{oj}^{\bf{X}} \times \left( {{\bf{g}}_{oj}^{\bf{X}}\cdot\nabla } \right)\nabla  \times {{\bf{A}}_{p1}}\\
 \approx  - \left( {{\bf{g}}_{oj}^{\bf{X}}\cdot\nabla } \right)\left( {{\bf{g}}_{oj}^{\bf{X}} \times \nabla  \times {{\bf{A}}_{p1}}} \right).
\end{array}
\end{equation}
To derive Eq.(\ref{af34}), the following two properties are used: equality $\nabla\cdot \mathbf{B}(\mathbf{X}_{oj},t)=0$; FLR term ${\left( {{\bf{g}}_{oj}^{\bf{X}}\cdot\nabla } \right){{\bf{B}}_{p1}}}$ is of the order lower than other non-FLR terms. Noting that ${{{\bf{A}}_{p1}}\left( {{{\bf{X}}_{oj}},t} \right)}$ is mainly contributed by the parallel part in Eq.(\ref{af22}), and the spatial gradient of ${{{\bf{A}}_{p1}}\left( {{{\bf{X}}_{oj}},t} \right)}$ are mostly contributed by the perpendicular part, the following equation can be derived
\begin{equation}\label{af35}
\begin{array}{l}
{\bf{g}}_{oj}^{\bf{X}} \times \nabla  \times {{\bf{A}}_{p1}}\\
 \approx {\bf{g}}_{oj}^{\bf{X}} \times \left( {{{\bf{e}}_ \bot }{\nabla _ \bot } \times {A_{p1\parallel }}{\bf{b}}} \right)\\
 = {\bf{g}}_{oj}^{\bf{X}} \times \left( {{{\bf{e}}_ \bot } \times {\bf{b}}} \right){\nabla _ \bot }{A_{p1\parallel }}\\
 = \left( { - {\bf{g}}_{oj}^{\bf{X}}\cdot{{\bf{e}}_ \bot }} \right){\bf{b}}{\nabla _ \bot }{A_{p1\parallel }}\\
 \approx  - \left( {{\bf{g}}_{oj}^{\bf{X}}\cdot{{\bf{e}}_ \bot }{\nabla _ \bot }} \right){{\bf{A}}_{p1\parallel}}.
\end{array}
\end{equation}
Then, the last term in Eq.(\ref{af34}) can be simplified to be
\begin{equation}\label{af36}
\begin{array}{*{20}{l}}
{ - \left( {{\bf{g}}_{oj}^{\bf{X}}\cdot\nabla } \right)\left( {{\bf{g}}_{oj}^{\bf{X}} \times \nabla  \times {{\bf{A}}_{p1\parallel }}\left( {{{\bf{X}}_{oj}},t} \right)} \right)}\\
{ = {{\left( {{\bf{g}}_{oj}^{\bf{X}}\cdot\nabla } \right)}^2}{{\bf{A}}_{p1\parallel }}\left( {{{\bf{X}}_{oj}},t} \right){\bf{b}},}
\end{array}
\end{equation}
which is the FLR term
Based on the homogeneous assumption, the summation of the first term of Eq.(\ref{af33})  is derived finally
\begin{equation}\label{af37}
\begin{array}{l}
\sum\limits_{o \in \{ i,e\} } {\sum\limits_j {{\varepsilon _o}L_{{\bf{g}}_{oj}^{\bf{X}}}^2\left[ {{{\bf{A}}_{p1}}\left( {{{\bf{X}}_{oj}},t} \right)\cdot d{{\bf{X}}_{oj}}} \right]} } \\
 = \sum\limits_{o \in \{ i,e\} } {\sum\limits_j {{\varepsilon _o}{{\left( {{\bf{g}}_j^{\bf{X}}\cdot\nabla } \right)}^2}{A_{p1\parallel }}\left( {{{\bf{X}}_{oj}},t} \right){\bf{b}}} } \\
 = \sum\limits_{o \in \{ i,e\} } {\sum\limits_j {{\varepsilon _o}\rho _j^2\nabla _{oj \bot }^2{A_{p1\parallel }}\left( {{{\bf{X}}_{oj}},t} \right){\bf{b}}} }.
\end{array}
\end{equation}

Eventually, $\Gamma_2$ in Eq.(\ref{af31}) is simplified to be
\begin{equation}\label{af38}
{\Gamma _2} = \sum\limits_{o \in \{ i,e\} } {\sum\limits_j {\left[ {{\varepsilon _o}\rho _j^2\nabla _{oj \bot }^2{A_{p1\parallel }}\left( {{{\bf{X}}_{oj}},t} \right){\bf{b}}\cdot d{{\bf{X}}_{oj}} - {\varepsilon _o}\frac{{2{m_o}{\mu _{oj}}}}{{{q_0}}}d{\theta _{oj}}} \right]} }.
\end{equation}

\subsection{The fundamental Lagrangian 1-form up to the second order approximation}

The fundamental Lagrangian 1-form up to the second order approximation is derived by summing $\Gamma_0, \Gamma_1, \Gamma_2$ together given by Eqs.(\ref{af16},\ref{af30},\ref{af38}), respectively
\begin{equation}\label{af39}
{\bar \Gamma _2} = \sum\limits_{o \in \{ i,e\} } {\sum\limits_j {\left[ {\begin{array}{*{20}{l}}
{\left( {\begin{array}{*{20}{l}}
{{q_o}\left( {{{\bf{A}}_E}\left( {{{\bf{X}}_{oj}}} \right) + {{\bf{A}}_{p0}}\left( {{{\bf{X}}_{oj}}} \right) + {{\bf{A}}_{p1}}\left( {{{\bf{X}}_{oj}},t} \right)} \right)}\\
{ + \rho _j^2\nabla _{oj \bot }^2{A_{p1\parallel }}\left( {{{\bf{X}}_{oj}},t} \right){\bf{b}} + {m_o}{U_{oj}}{\bf{b}} + {m_o}{{\bf{v}}_{ojd}}}
\end{array}} \right)\cdot d{{\bf{X}}_{oj}}}\\
{ + \frac{{{m_o}{\mu _{oj}}}}{{{q_o}}}d{\theta _{oj}} - \left( {{\mu _o}B\left( {{{\bf{X}}_{oj}}} \right) + \frac{{{m_o}U_{oj}^2}}{2} + \frac{{{m_o}v_{ojd}^2}}{2}} \right)dt}
\end{array}} \right]} } .
\end{equation}
It should be noted that the FLR term for each particle is introduced in this fundamental Lagrangian 1-form on the new coordinates. The term $\mathbf{v}_{ojd}$ is just given by the drift velocity of the charged particle.

\newpage
\section*{References}

\bibliographystyle{pst}
\bibliography{GAP}

\end{document}